# Parallel photonic accelerator for decision making using optical spatiotemporal chaos

Kensei Morijiri[1,*], Kento Takehana[1], Takatomo Mihana[1,2], Kazutaka Kanno[1], Makoto Naruse[2], and Atsushi Uchida[1,**]

[1] *Department of Information and Computer Sciences, Saitama University, 255 Shimo-okubo, Sakura-ku, Saitama City, Saitama 338-8570, Japan*

[2] *Department of Information Physics and Computing, Graduate School of Information Science and Technology, The University of Tokyo, 7-3-1 Hongo, Bunkyo-ku, Tokyo 113-8656, Japan*

Corresponding authors' emails: [*]kensei.1221.0926.snow@gmail.com, [**]auchida@mail.saitama-u.ac.jp

**Abstract**

Photonic accelerators have attracted increasing attention in artificial intelligence applications. The multi-armed bandit problem is a fundamental problem of decision making using reinforcement learning. However, the scalability of photonic decision making has not yet been demonstrated in experiments, owing to technical difficulties in physical realization. We propose a parallel photonic decision-making system for solving large-scale multi-armed bandit problems using optical spatiotemporal chaos. We solve a 512-armed bandit problem online, which is much larger than previous experiments by two orders of magnitude. The scaling property for correct decision making is examined as a function of the number of slot machines, evaluated as an exponent of 0.86. This exponent is smaller than that in previous work, indicating the superiority of the proposed parallel principle. This experimental demonstration facilitates photonic decision making to solve large-scale multi-armed bandit problems for future photonic accelerators.

Photonics approaches to information processing have attracted increasing attention to accommodate the demands of machine learning and artificial intelligence, which in part stem from the fundamental limits of conventional semiconductor integration technologies, as described by the end of Moore's law [1–4]. Photonic technologies based on time, space, and wavelength multiplexing can realize fast and massive parallel implementations of machine learning schemes to overcome the limitation of information-processing speed in recent semiconductor technologies. High-speed and energy-efficient information processing is indispensable for cyber-physical systems and digital societies based on big data. Recently, photonic accelerators have been intensively studied to enhance specialized tasks in machine learning [5–26], and the unique physical attributes of photons (e.g., multiple degrees of freedom with intensity, wavelength, and polarization) have been exploited for advanced photonic technologies in optical communication and photonic circuit integration [5]. Examples of photonic accelerators are photonic neural networks [6,7], coherent Ising machines [8], optical pass gate logic [9], photonic reservoir computing [10–13], and photonic decision making [14–26].

The multi-armed bandit (MAB) problem has been investigated extensively in photonic decision making; it is a fundamental problem in reinforcement learning. The MAB problem aims to maximize the total reward with a finite number of plays by selecting one among multiple slot machines, called arms, with initially unknown hit probabilities. First, the player randomly selects slot machines to search for the slot machine with the highest hit probability (called exploration). The player then extensively selects the slot machine estimated to be the best for maximizing the total reward (called exploitation). There is a trade-off known as the exploration–exploitation dilemma [27,28]. Significant exploration leads to the correct estimation of the best slot machine. However, the total reward is reduced owing to the lack of exploitation. By contrast, high exploitation increases the total reward if the best slot machine is correctly estimated. However, the estimation may not be correct owing to insufficient exploration. The MAB problem is directly associated with several practical applications, for example, channel selection in wireless and optical communication networks [29–31].

Several schemes have been proposed for photonic decision making by leveraging the attributes of the fast, irregular, and complex dynamics of photons. For example, two-armed bandit problems involving two slot machines have been successfully resolved using single photons [14,15], the chaotic temporal waveforms of laser output intensities [16–19], mode competition dynamics [20,21], and synchronization phenomena in coupled semiconductor lasers [22–25]. Some of these schemes have been extended to solve the MAB problem with a large number of slot machines using hierarchical structures [15,17], coupled laser networks [23], multi-mode laser dynamics [21], and the bias control of chaotic temporal waveforms [26].

These principles have been examined extensively using theoretical analysis and numerical simulations. Experimental studies have been limited to solving MAB problems with two [14,16,20,22], three [25], and four [15] slot machines owing to the experimental and technological difficulties in extending to a large number of slot machines. In these previous studies, bulky optical components have been used to represent slot machines in experimental implementations, and the extension of these systems to a large number of slot machines is not straightforward. In addition, good scalability cannot be achieved in terms of the number of slot machines [17,23].

Recently, chaotic spatiotemporal dynamics have been utilized for reservoir computing, which is a simplified version of recurrent neural networks [32–34]. Chaotic spatiotemporal dynamics have been generated in an optoelectronic feedback system with a spatial light modulator (SLM) and camera (e.g., fast vision chip), and the pixels on the SLM are considered to be the spatial optical neurons used for reservoir computing. The implementation of reservoir computing in solving parallel tasks can be achieved using a large number of spatial optical neurons in these schemes. This technique has also been used to implement photonic spiking neural networks [35]. Such a spatiotemporal scheme could be a promising resource to experimentally realize decision making with a large number of slot machines (over 100).

In this study, we experimentally demonstrate a massively parallel architecture of photonic decision making for solving a large-scale MAB problem using optical spatiotemporal chaos as a photonic accelerator. The optical spatiotemporal chaos is generated in an optoelectronic feedback system with an SLM, a camera, and a signal processing module. We experimentally solve MAB problems for up to 512 slot machines, which is beyond the maximum number of slot machines (four) experimentally demonstrated in the literature. We investigate the scaling characteristics of the proposed method, which outperforms existing software-based algorithms.

**Results**

*Spatiotemporal dynamics in an optoelectronic feedback system*

Fig. 1 shows the experimental setup of the parallel architecture for decision making. We used an optoelectronic feedback system comprising a semiconductor laser, SLM, complementary metal-oxide semiconductor (CMOS) camera, and personal computer (PC). The laser beam was expanded to generate a two-dimensional (2D) beam pattern using a microscope objective and was subjected to the SLM. The optical phase of the 2D beam pattern was modulated by the SLM, and phase modulation was converted into intensity modulation by two polarizers. The intensity-modulated 2D beam pattern was measured using the camera. The optical intensities were processed in the PC to generate the subsequent phase-modulation signal to be applied in the SLM. This procedure was repeated, and 2D spatiotemporal dynamics were accomplished in this optoelectronic feedback system. Further details of the experimental setup are described in the *Methods* section.

The update formula of the 2D pattern displayed in the SLM, or equivalently, the input-output characteristics of the CMOS camera in the optoelectronic feedback system, is described as the following discrete map:

$$S_i^{CAM}(t+1) = a \cdot \cos\left(2\pi f \beta S_i^{CAM}(t)\right) + b, \quad (1)$$

where $S_i^{CAM}(t)$ is the average optical intensity of macro-pixel *i* at time *t* detected by the camera. A macro-pixel comprises a group of neighboring pixels on the SLM and camera, as defined in the *Methods* section. Experimentally, $S_i^{CAM}(t)$ is detected as a value with 12-bit resolution, which is then converted into an 8-bit signal to match the resolution of the SLM (we discard the four least significant bits). In fact, this system can be interpreted as the Ikeda map [36,37]. In Eq. (1), *a*, *b*, and *f* are parameters corresponding to the amplitude, offset, and frequency of the sinusoidal map, respectively. $\beta$ is the feedback coefficient that determines the number of local maxima and minima in the sinusoidal map and is the bifurcation parameter that determines spatiotemporal dynamics. In the experiment, $\beta =$ 3.2 is used to generate chaotic spatiotemporal dynamics. The derivation of Eq. (1) as well as its numerical results are provided in the *Supplementary Information*.

Fig. 2(a) shows a snapshot of the spatiotemporal dynamics detected using the camera. An irregular spatial pattern of 8 × 8 macro-pixels is observed because each macro-pixel begins with a different initial condition, leading to the observation of versatile dynamics at each macro-pixel. Fig. 2(b) shows an example of the temporal dynamics of the laser intensity at macro-pixel (4, 4) (located in the fourth row and fourth column). The temporal dynamics exhibit chaotic fluctuations when the feedback strength is large enough, as normally observed in the Ikeda map. Independent chaotic oscillations are observed at different macro-pixels because of the sensitive dependence on initial conditions. A movie of the spatiotemporal dynamics is provided in the *Supplementary Information*.

Fig. 2(c) shows the one-dimensional map generated in the optoelectronic feedback system at macro-pixel (4, 4) when the feedback coefficient is set to $\beta = 3.2$. The absolute values of the derivatives of the map at the crossing points are larger than 1, indicating the generation of chaotic dynamics. The parameter values are estimated from Fig. 2(c) as follows: $a = 101$, $b = 104$, and $f = 1/201$. Each nonlinear function is different for each macro-pixel, and therefore the nonlinear function must be adjusted by changing the $a$, $b$, and $f$ values of the sinusoidal map, as described in the *Methods* section. Fig. 2(d) shows the histogram of the chaotic temporal waveform in Fig. 2(b). Double peaks are observed at both edges of the histogram, which originate from the characteristics of the sinusoidal map, as shown in Fig. 2(c). Such a double-peak distribution is advantageous for efficient decision making, as discussed in the following section.

*Decision-making results*

In the proposed decision-making method for solving MAB problems, we first assign each slot machine to each macro-pixel on the SLM to select a slot machine based on the optical intensities. We introduce a bias in the optical intensity for decision making [26], and measure the optical intensity of the light beams modulated via the SLM using the camera, which is denoted by $I_i(t)$ for macro-pixel $i$ at time $t$. The optical intensity is biased using the following formula:

$$A_i(t) = I_i(t) + kB_i(t), \qquad (2)$$

where $B_i(t)$ denotes the bias given to macro-pixel $i$ at time $t$, and $k$ is the bias coefficient. The biased intensity $A_i(t)$ is compared among all macro-pixels, and the maximum $A_i(t)$ is determined. At time $t$, the decision is to select slot machine $i$ according to the maximum $A_i(t)$. After playing slot machine $i$, if the result of slot machine $i$ is a hit, the corresponding bias $B_i(t)$ is increased, and other biases $B_j(t)$ ($j \neq i$) are decreased. Thus, slot machine $i$ is highly likely to be selected in the subsequent selections. By contrast, if the result of slot machine $i$ is a miss, the corresponding bias $B_i(t)$ is decreased, and other biases $B_j(t)$ ($j \neq i$) are increased; therefore, slot machine $i$ is selected less frequently. The bias $B_i(t)$ is determined based on the result of slot machine selection using the tug-of-war method [26,38–40], as summarized in the *Methods* section.

After the detection of optical intensities by the camera, signal processing, in the decision-making process, is conducted electrically in the PC, ranging from estimating hit probabilities, calculating biases, slot machine selection, emulating slot machines, and generation of the phase modulation signal to the SLM. The entire decision-making process is executed online in the optoelectronic feedback system. The potential technological advancements for parallel computation in postprocessing are discussed in the *Discussion* section.

First, we demonstrate the successful solution of a 64-armed bandit problem based on the proposed parallel photonic decision making. The hit probabilities of the slot machines or arms are configured as follows: $P_1 = 0.7$, $P_2 = 0.5$, $P_3 = 0.9$, $P_4 = 0.1$, …, $P_{2j-1} = 0.7$, and $P_{2j} = 0.5$, $j \geq 3$, where $j$ is an integer [17,26]. In this case, slot machine 3 has the highest hit probability, and the correct decision is to select slot machine 3. Fig. 3 shows the experimental results of decision making to solve the 64-armed bandit problem. Figs. 3(a) and 3(b) present the cross-sections of the spatiotemporal patterns of the chaotic temporal waveforms with biases at the beginning (1st play) and final play (1000th play), respectively. It can be observed that the macro-pixel (1, 3), which corresponds to slot machine 3, exhibits a large value, indicated by the white color at the 1000th play in Fig. 3(b), whereas the other macro-pixels show smaller signal levels indicated by the darker colors. This indicates that the correct decision has been successfully made, meaning that slot machine 3 was frequently selected. A movie of the decision-making process on the spatiotemporal dynamics of optical intensity with bias is provided in the *Supplementary Information* to demonstrate the behavior of the proposed decision-making method.

The time evolution of the chaotic signal level with the biases shown in Fig. 3(c) reveals that the signal level of slot machine 3 exhibits dramatic increases after approximately the 400th play, whereas the other slot machines (we only show slot machines 1, 2, and 4 for simplicity) continue to exhibit lower signal levels, with a maximum amplitude of approximately 200. Similarly, Fig. 3(d) shows the evolution of the selected slot machine index. In the early phase, it can be seen that a variety of slot machines are randomly selected, whereas slot machine 3 is selected more frequently up to approximately the 500th play, followed by the selection of slot machine 3 only. Therefore, decision making has been performed adequately.

### *Statistical evaluation of decision-making performance*

In this section, we investigate the statistical characteristics of the decision-making performance. We introduce the correct decision rate (CDR), which is described as follows [16]:

$$CDR(t) = \frac{1}{n}\sum_{i=1}^{n} C(i, t), \tag{3}$$

where $C(i, t)$ represents a function that returns 1 if the highest-hit-probability slot machine is selected for the $t$-th play ($t = 1, 2, \cdots, m$) and $i$-th cycle, otherwise it returns 0. $m$ denotes the number of plays, and $n$ denotes the number of cycles. A large CDR implies that the highest-hit-probability slot machine is selected more often. The decision-making process was repeated for at least 100 cycles to statistically evaluate the decision-making performance. The number of plays was changed for different $N$ until the CDR converged to 0.95.

Fig. 4(a) summarizes the evolution of the CDR for different numbers of slot machines $N$, ranging from 8 to 512. The CDR for all $N$ curves increases monotonically after the initial exploration duration and exceeds 0.95. Therefore, correct decision making is performed even when the number of slot machines increases to $N = 512$. Furthermore, we examine the number of plays when the CDR exceeds 0.95 as a function of the number of slot machines, which is indicated by the black curve in Fig. 4(b). Here, the relationship between the number of plays $y$ required to reach a CDR of 0.95 and the number of slot machines $N$ can be approximated by a power law: $y = 30.0\, N^{0.86}$. Therefore, the scaling exponent is 0.86, which is less than 1. This indicates that the scaling exponent is smaller than those in previous reports (e.g., 1.16 in [17] and 1.85 in [23]). Therefore, the proposed method is advantageous when the number of slot machines is large.

This superior scaling characteristic of the proposed method could be caused by the amplitude distribution of the chaotic waveforms. The probability distribution at small and large amplitudes (near 0 and 192) is larger than those at other signal levels, as shown in Fig. 2(d). The best machine is associated with a large bias value, and therefore a high probability of a large-amplitude chaotic signal will lead to a high likelihood of yielding the maximum $A_i(t)$ in Eq. (2), which provides the decision to choose slot machine $i$. In this manner, the peak of the large-amplitude chaotic signal can accelerate the exploitation of the best slot machine in the present architecture. Simultaneously, the peak of the small-amplitude chaotic signal probability implies that minute differences affect $A_i(t)$ in Eq. (2) in the early stages, implying that the system accumulates sufficient exploration. A deeper understanding of the underlying principle and optimization of chaotic dynamics are interesting directions for future studies.

*Performance comparison with other algorithms*

We compare the performance of the proposed method with those of other algorithms. We use a previously reported decision-making method using chaotic temporal waveforms generated by semiconductor lasers [26]. In addition, we use Thompson sampling [41] and the upper confidence bound 1-tuned (UCB-1 tuned) algorithm [42], which are well-known for solving the MAB problem.

Fig. 5 compares the scaling characteristics of the proposed parallel photonics method using the SLM experiment, laser-chaos-based method, Thompson sampling, and UCB-1 tuned algorithm. We calculate the CDR for different $N$, and compare the scaling characteristics of these methods. Fig. 5 shows that the proposed method using the SLM experiment needs the smallest number of plays to reach a CDR of 0.95 for different $N$. The scaling exponents are 0.86, 0.98, 1.13, and 1.08 for the proposed method, laser-chaos-based method, Thompson sampling, and UCB-1 tuned algorithm, respectively. The scaling exponent for the proposed method is the smallest, and hence the proposed method outperforms the other decision-making methods. In fact, the number of plays needed to achieve a CDR of 0.95 for the proposed method is 6.5 times smaller than that for the UCB-1 tuned algorithm for $N = 512$. The curves for the proposed method and laser-chaos-based method are approximately similar. However, the curve for the SLM experiment is slightly smaller than that for the laser-chaos-based method for a large number of slot machines (over 100). Moreover, the number of plays needed to reach a CDR of 0.95 for Thompson sampling and the UCB-1 tuned algorithm is larger than those for the proposed and laser-chaos-based methods. Therefore, the scaling characteristics of the proposed method are superior to those of other methods. Another statistical measure, *regret,* is described in the *Methods* section.

**Discussion**

We experimentally constructed a parallel photonic decision-making system using chaotic spatiotemporal dynamics with an SLM and electrical processing, which successfully solves the 512-armed bandit problem. All decision-making processes are performed online in an automatically controlled optoelectronic feedback system. To the best of our knowledge, this is the first online experimental demonstration of decision making for such a large number of slot machines (up to 512). It is worth noting that with the current experimental system, the number of slot machines can be expanded to the order of $10^5$ in a straightforward manner, as the maximum macro-pixels can accommodate a maximum of 262,144 ($512 \times 512$) slot machines.

The execution time for one play in the decision-making process is 360 ms, including the detection of an optical signal by the camera, signal processing for decision making in the PC, and feedback modulation on the SLM. The primal latency is derived from the software processing performed by the CPU in the PC. However, the priority of the current experimental implementation is the execution of the entire system, rather than an exploration of the ultimate technological possibilities. This software processing can be replaced by specialized hardware such as a field-programmable gate array (FPGA) for faster decision making. In addition, the frame rates of spatial light modulation and imaging can be improved by using faster equipment up to the kilohertz range, in view of recent advances in digital micromirror devices [43] and high-speed and massive resolution CMOS image sensors [44,45]. We note that signal processing for chaotic transformation using Eq. (1) is extremely simple, and complete pixel-parallel processing is possible. Therefore, merging photodetection and signal processing into a single-pixel level, which is sometimes referred to as a vision chip or smart pixels in the literature [46,47], could be a promising approach for parallel photonic decision making.

Parallel architecture is one of the advantages of spatiotemporal feedback systems. Currently, a single execution rate of one play in a photonic system is an order of magnitude slower than it is in electrical software processing such as Thompson sampling. However, it should be noted that the parallel approach could outperform Thompson sampling and the UCB-1 tuned algorithm even in the current setup because of the parallel nature of the proposed system. For example, when the number of slot machines $N$ is 1024 or more and the number of cycles is 512 or more, the total execution time of the proposed method will be less than the time for Thompson sampling, even under the current experimental conditions. This can be attributed to the parallel nature of the current optoelectronic feedback systems, where the number of slot machines is determined by the number of macro-pixels on the SLM. This approach can be easily extended to the spatial degree of freedom, and spatiotemporal systems can be a promising resource for large-scale decision making.

In the current experiment, coupling among macro-pixels is avoided for the purpose of completely independent spatiotemporal chaos among the macro-pixels. More complex spatiotemporal dynamics with correlations will be observed when inter-macro-pixel coupling exists, which could lead to more efficient decision making for a larger number of slot machines. Moreover, such coupling is easily optically achievable, as demonstrated in the latest computational imaging techniques [48]. Furthermore, the synchronization of spatiotemporal chaos and chimera states can be observed in optoelectronic feedback systems [49,50]. The introduction of coupling among macro-pixels and the tuning of spatiotemporal dynamics for efficient decision making would be interesting future work.

In this study, we experimentally demonstrated a parallel photonic decision-making system for solving large-scale MAB problems using optical spatiotemporal chaos generated by an SLM and CMOS imaging camera. We generated spatiotemporal dynamics in an optoelectronic feedback system using the SLM and camera. By associating macro-pixels on the SLM to the slot machines in the MAB problem, a parallel architecture was successfully implemented with up to 512 slot machines. The study was conducted on a completely online and automated experimental apparatus, which was 128 times larger than previous experiments that addressed four-armed bandit problems in the literature. Furthermore, we examined the scaling characteristic of decision-making performance as a function of the number of slot machines, where a power-law relationship with a scaling exponent of 0.86 was found, which was smaller than those reported for previous methods, including the well-known software-based bandit algorithms. Although the primary latency stems from the CPU computations in the PC, the parallel architecture is matched with the latest FPGA or even photodetection-and-processor-integrated vision chips for enhanced acceleration. Our results demonstrate the parallel properties of light and the advantages of parallel photonics technologies to higher-order functionalities, such as decision making, reinforcement learning, and artificial intelligence.

## Methods
### *Experimental setup*
The details of the experimental setup are shown in Fig. 1. A distributed feedback semiconductor laser (Thorlabs, LP642-PF20) with a 642 nm light source was collimated using a microscope objective, and the size of the laser beam was expanded to generate a 2D beam pattern. The laser power was adjusted using a neutral-density filter. A spatial filter was used to generate a flat wavefront, and the laser light was linearly polarized by a polarizer and sent to an SLM (X13138-01, Hamamatsu photonics, 1272 × 1024 pixels, 15.9×12.8 mm effective size, 60 frame/s). The phase of the reflected light was modulated by the SLM, and the conversion from phase into intensity modulation was accomplished using another polarizer. The laser light was detected by a CMOS camera (C11440-36U, Hamamatsu photonics, 1920

× 1200 pixels, 11.25 × 7.03 mm effective size, 64.9 frame/s), and the 2D pattern of the laser intensity was recorded in a PC (Dell, CPU: Intel Core i7-9700, 3.00 GHz, RAM: 8.0 GB, OS: Windows 10). Decision making was performed in the PC based on the 2D optical intensities detected at the macro-pixels, where slot machines are emulated in the PC. After one play of a selected slot machine for decision making, a phase modulation signal to the SLM was generated based on the detected optical intensities and calculated biases, and fed back to the macro-pixels of the SLM. The mismatch between the pixel sizes of the camera and SLM was adjusted in the PC. Therefore, the optoelectronic feedback loop was implemented using an SLM, CMOS camera, and PC to generate spatiotemporal chaos as well as to perform decision making. This procedure was repeated until the final play was completed for decision making.

### *Definition of macro-pixels*

We define the notion of a macro-pixel in an SLM and a CMOS camera. A macro-pixel comprises a group of neighboring pixels on the SLM and CMOS camera. We assume that $R \times R$ consecutive pixels on the SLM correspond to an individual slot machine, and a macro-pixel refers to the individual block occupying $R \times R$ pixels. The optical intensity is averaged over the macro-pixel and considered as $S_i^{SLM}$. The average intensity $S_i^{SLM}$ of the macro-pixel on the SLM is converted into that of the CMOS camera $S_i^{CAM}$ by magnifying the 2D beam pattern.

When the total number of pixels of the SLM is given by $S \times S$, the total number of macro-pixels is given by $N = M \times M$, where $M = S / R$. $S$ is specified as 512 in the experimental SLM used in the present study. The macro-pixel at the $r$-th row and $s$-th column ($r, s = 1, 2, …, M$) on the SLM is denoted as macro-pixel $i$, where $i = (r - 1)M + s$, which corresponds to slot machine $i$ ($i = 1, 2, …, N$). When the number of slot machines is smaller than $M^2$, some macro-pixels are not utilized in the system. In Figs. 2 and 3, $R$ is set to 64. In this case, there are 8 × 8 macro-pixels on the SLM, which accommodates a bandit problem with up to 64 arms (i.e., $M = 8$ and $N = 64$).

To avoid coupling among the macro-pixels, we eliminate the optical intensities around the edges of the macro-pixels. For instance, we average the intensities from the 54 × 54 pixels in the center of a macro-pixel (64 × 64 pixels), and the average intensity value is fed back to the corresponding macro-pixel on the SLM. In other words, we do not incorporate spatial couplings among the macro-pixels. Spatial coupling may induce more complex spatiotemporal dynamics, which will be investigated in future studies.

To adjust the size of macro-pixels between the SLM and camera, we convert the size of the macro-pixels in the PC. Specifically, the size of the macro-pixels on the SLM is magnified 2.25 times to adjust the size of the macro-pixels on the camera (i.e., 512 × 512 pixels on the SLM and 1152 × 1152 pixels on the camera are matched).

### *Algorithm for the calculation of bias using the tug-of-war method*

The bias, $B_i(t)$, in Eq. (2) is determined based on the result of slot machine selection by the tug-of-war method, which is expressed as follows [26,38–40]:

$$B_i(t) = Q_i(t) - \frac{1}{N-1}\sum_{i' \neq i}^{N} Q_{i'}(t), \quad (7)$$

$$Q_i(t) = \Delta W_i - \omega L_i, \quad (8)$$

$$\Delta = 2 - (\hat{P}_{top1} + \hat{P}_{top2}), \quad (9)$$

$$\omega = \hat{P}_{top1} + \hat{P}_{top2}, \quad (10)$$

$$\hat{P}_i = \frac{W_i}{T_i}, \quad (11)$$

where $Q_i(t)$ is the evaluation value of slot machine $i$, and $N$ is the number of slot machines. $T_i$, $W_i$, and $L_i$ are the number of total, hit (win), and miss (lose) selections for slot machine $i$, respectively. $\Delta$ and $\omega$ denote the coefficients for the hit and miss selections in the tug-of-war method, respectively. $\hat{P}_i$ denotes the estimated hit probability of slot machine $i$. $\hat{P}_{top1}$ and $\hat{P}_{top2}$ are the highest and second-highest estimated hit probabilities, respectively. The algorithm is slightly modified by introducing $\Delta$ and $\omega$ to achieve correct decision making for different settings of hit probabilities [22,26]. In particular, the proposed algorithm works properly even if both $\hat{P}_{top1}$ and $\hat{P}_{top2}$ are close to 0 or 1, whereas $\omega = (\hat{P}_{top1} + \hat{P}_{top2})/(2 - (\hat{P}_{top1} + \hat{P}_{top2}))$ in [26] becomes too small or large in this condition.

### Evaluation of regret

We introduce *regret* to evaluate the statistical characteristics of decision making, which is defined as follows:

$$Regret(p) = p\, P_{max} - \frac{1}{S}\sum_{l=1}^{S}\sum_{m=1}^{M}\left(P_m\, S_{l,m}(p)\right), \quad (12)$$

where $p$ is the number of plays, $P_m$ is the hit probability of slot machine $m$, $P_{max}$ is the maximum hit probability, $S$ is the total number of cycles, and $S_{l,m}(p)$ is the number of selections for slot machine $m$ at the $l$-th cycle until the $p$-th play. A smaller regret implies better decision-making performance.

Fig. 6(a) shows the regret as the number of plays is changed for different number of slot machines $N$. Regret increases and saturates at a certain value for all the cases. Fig. 6(b) shows scalability characteristics evaluated using regret at the 6000[th] play for the proposed method and Thompson sampling. The scaling exponents are 0.94 and 1.11 for the proposed method and Thompson sampling, respectively. Therefore, we confirm that the proposed method outperforms Thompson sampling in terms of regret.

**Data availability**

The datasets generated during the current study are available from the corresponding author upon reasonable request.

**Acknowledgements**

This work was supported in part by the Grants-in-Aid for Scientific Research from the Japan Society for the Promotion of Science (JSPS KAKENHI, Grant No. JP19H00868, JP20K15185, JP20H00233), JST CREST, Japan (JPMJCR17N2), and the Telecommunications Advancement Foundation.

## Author contributions

K. M., K. K., and A. U. designed the system architecture and principles. K. M. and K. T. conducted the experiment on decision making. K. M., K. T., T. M., K. K., and A. U. analyzed the data. K. M., M. N. and A. U. wrote the manuscript.

## Competing interests

The authors declare no competing interests.

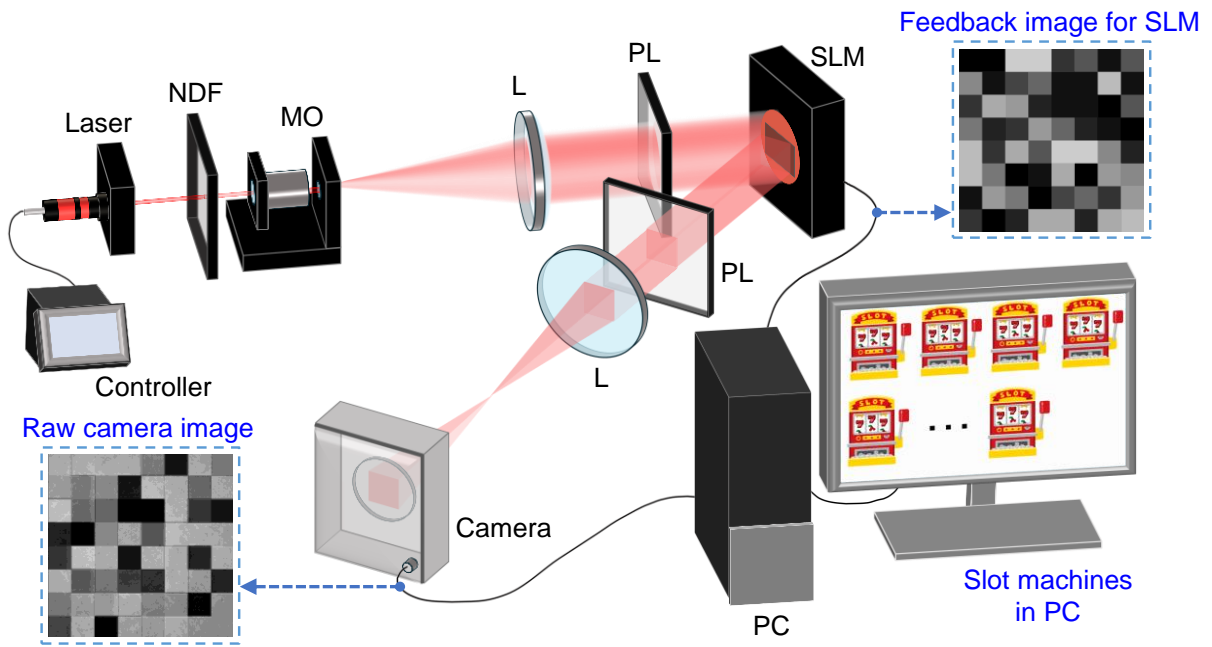

Figure 1. Experimental setup of an optoelectronic feedback system for decision making using optical spatiotemporal chaos. L, lens; MO, microscope objective; NDF, neutral-density filter; PC, personal computer; PL, polarizer; SF, spatial filter; SLM, spatial light modulator.

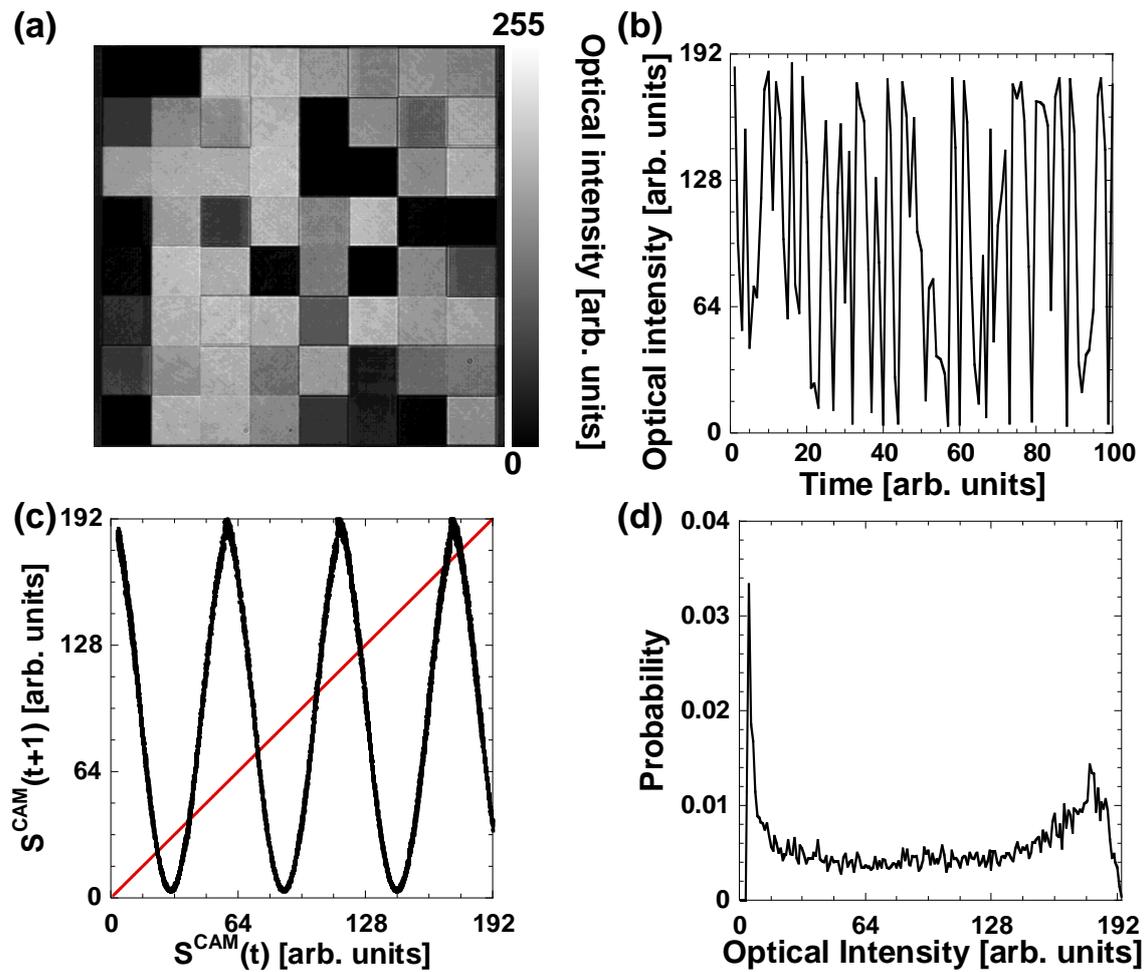

Figure 2. Experimental results of the dynamics of optical spatiotemporal chaos. (a) Spatiotemporal pattern of optical intensities on 8 × 8 macro-pixels. (b) Temporal dynamics of optical intensity reflected from the macro-pixel at the fourth row and fourth column. (c) Nonlinear function of the optoelectronic feedback system when the feedback coefficient is set to $\beta = 3.2$. Red line indicates the diagonal line. (d) Probability of the amplitude of chaotic temporal dynamics in (b).

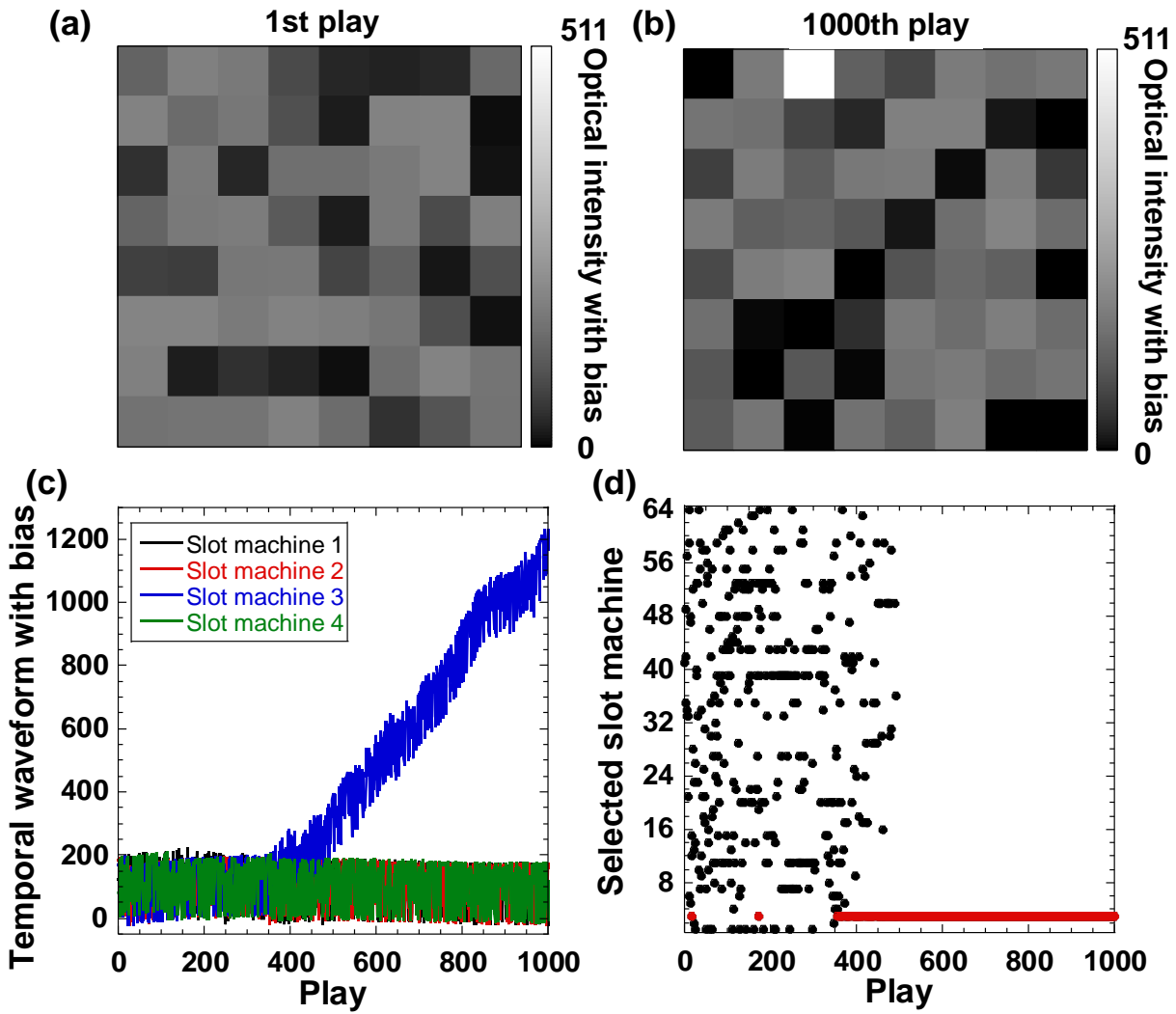

Figure 3. Experimental results of decision making for solving the 64-armed bandit problem. (a), (b) Spatiotemporal patterns of the optical intensities with biases for 8×8 macro-pixels at the (a) first and (b) final (1000th) plays. The averaged value is plotted for each macro-pixel. (c) Temporal waveforms of optical intensities with biases assigned to slot machines 1 to 4. (d) Selected slot machines as the number of plays is changed. Red dots indicate the correct selection of slot machine 3 with the highest hit probability.

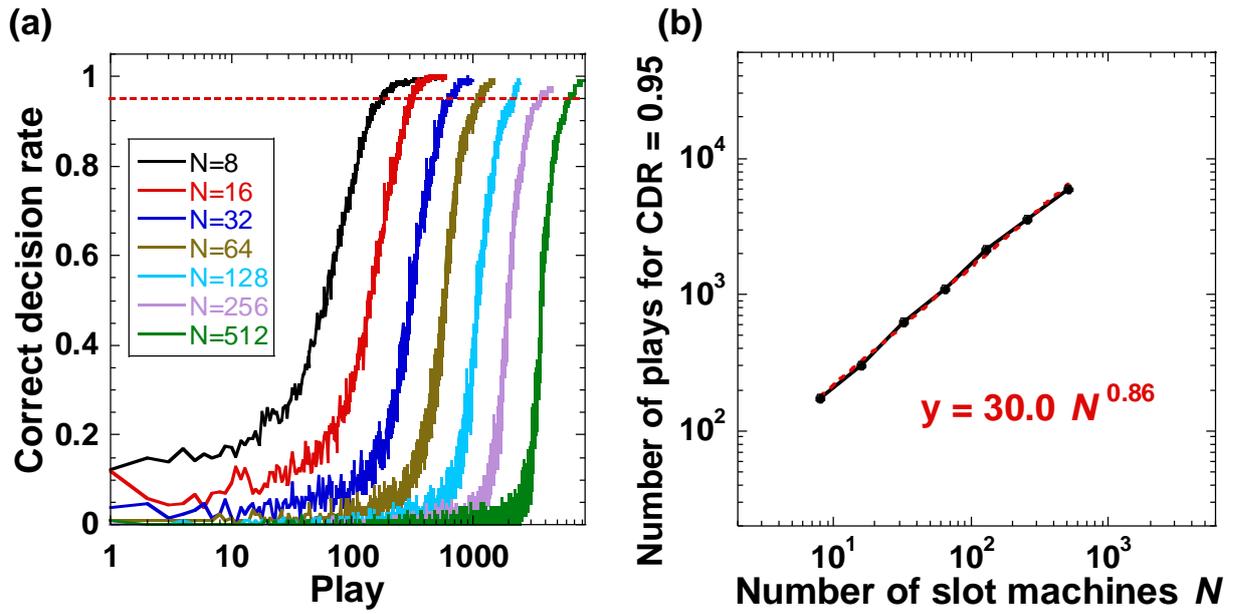

Figure 4. Experimental results of the statistical characteristics of decision making. (a) Correct decision rate (CDR) as the number of plays is changed for different numbers of slot machines from $N = 8$ to 512. The red dotted line indicates a CDR of 0.95. (b) Scaling characteristics (black) between the number of plays for CDR = 0.95 and the number of slot machines $N$. The data are approximated by a power law, indicated as the red dotted line. The bias coefficient $k = 15$ is used for different $N$ after optimization.

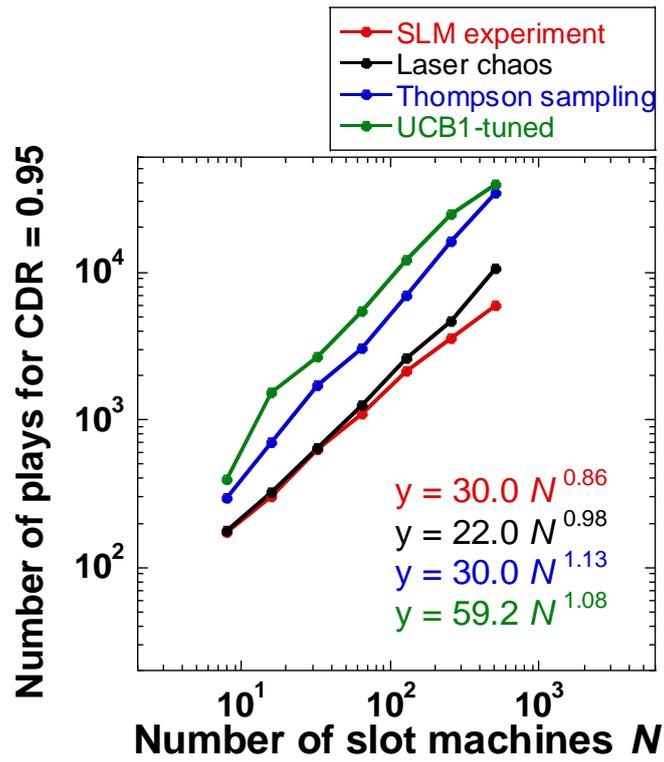

Figure 5. Comparison of scaling characteristics between the number of plays (*y*) needed for CDR = 0.95 and the number of slot machines *N*. Results for the proposed method using the SLM experiment (red), laser-chaos-based method (black), Thompson sampling (blue), and UCB-1 tuned algorithm (green) are shown.

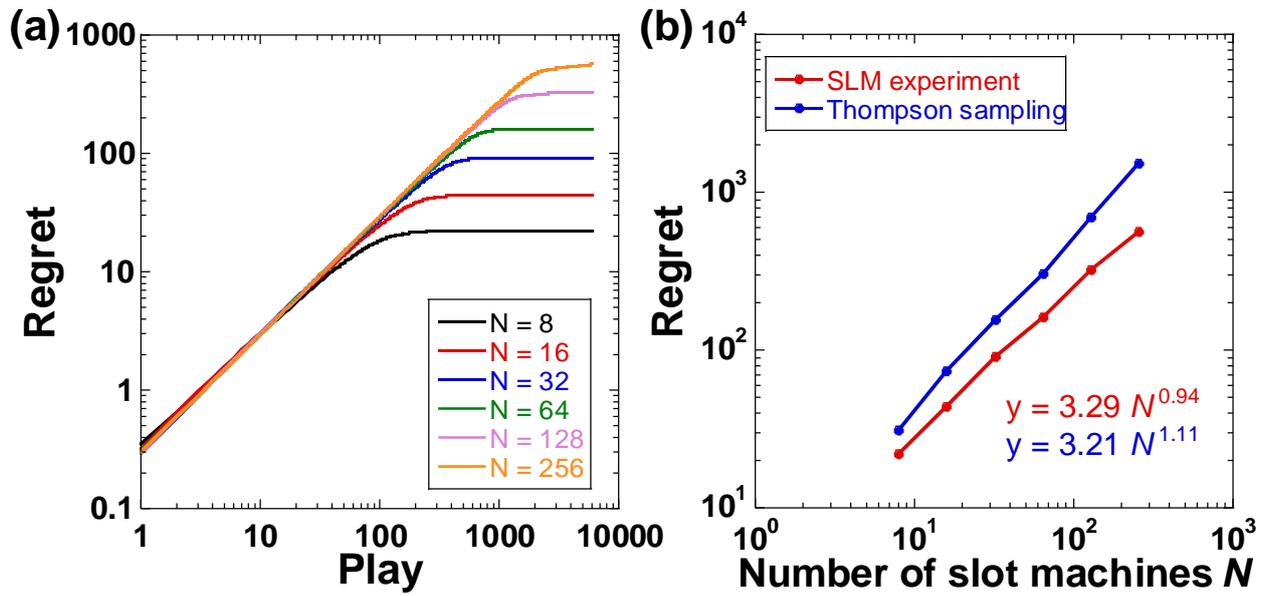

Figure 6. (a) Regret of the proposed method using the SLM experiment for different numbers of slot machines $N$. (b) Comparison of regret at the $6000^{th}$ play of the proposed method and Thompson sampling. Regret is evaluated as a function of the number of slot machines $N$.

# Supplementary Information of Parallel photonic accelerator for decision making using optical spatiotemporal chaos

Kensei Morijiri[1,*], Kento Takehana[1], Takatomo Mihana[1,2], Kazutaka Kanno[1], Makoto Naruse[2], and Atsushi Uchida[1,**]

[1] Department of Information and Computer Sciences, Saitama University, 255 Shimo-okubo, Sakura-ku, Saitama City, Saitama 338-8570, Japan

[2] Department of Information Physics and Computing, Graduate School of Information Science and Technology, The University of Tokyo, 7-3-1 Hongo, Bunkyo-ku, Tokyo 113-8656, Japan

Corresponding authors' emails: [*]kensei.1221.0926.snow@gmail.com, [**]auchida@mail.saitama-u.ac.jp

**Numerical simulation and results**

Numerical simulations were performed using the chaotic map in Eq. (1), as indicated in the main text. The parameter values for the numerical simulations were set as follows: $a = 101$, $b = 104$, and $f = 1/201$, obtained from the experimental result in Fig. 2(c) in the main text. $\beta$ is the bifurcation parameter that determines the spatiotemporal dynamics.

Fig. S1(a) shows an example of the chaotic temporal waveform obtained from numerical simulations. Chaotic irregular behavior is observed in Fig. S1(a). Fig. S1(b) shows the probability of the amplitude of the chaotic temporal waveform. Two peaks appear at the edges of the intensity. Fig. S1(c) shows the nonlinear function of the chaotic map for the feedback coefficient $\beta = 3.2$. The nonlinear function obtained from the numerical simulation matches that obtained by the experiment, as shown in Fig. 2(c) in the main text. These numerical results agree well with the experimental results shown in Fig. 2.

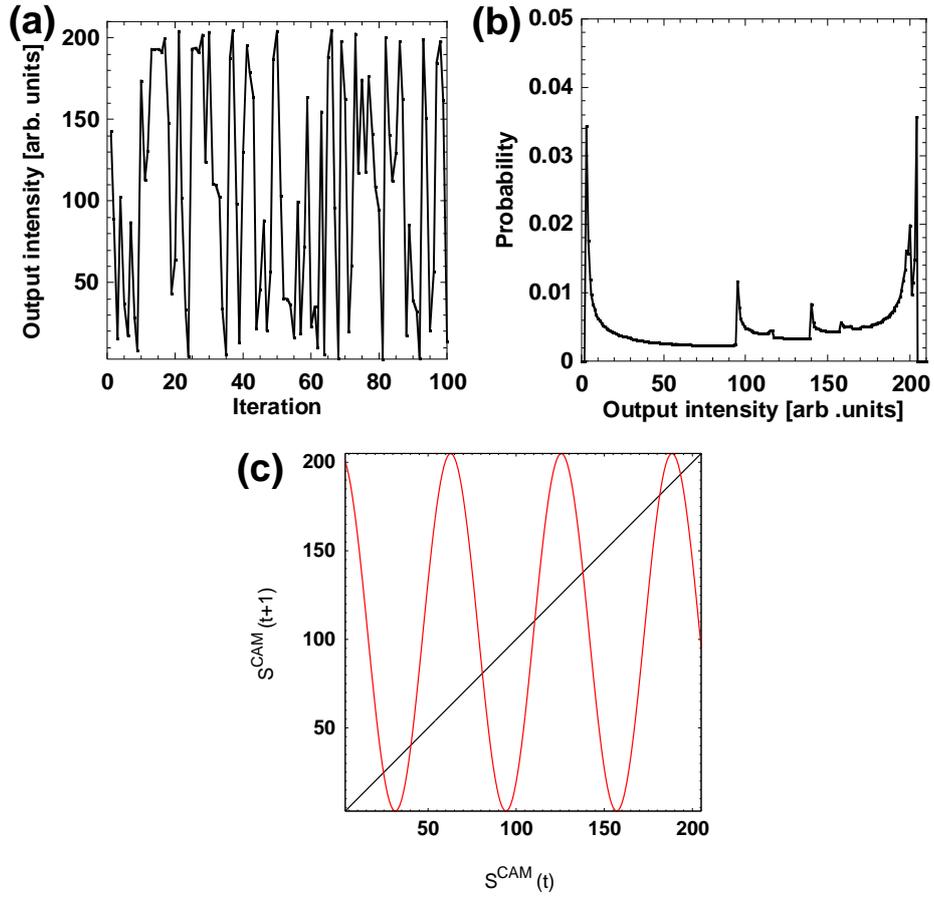

Figure S1. Numerical results of the chaotic map. (a) Temporal waveform, (b) probability of the amplitude of the temporal waveform, and (c) nonlinear function of the chaotic map for $\beta = 3.2$.

We investigated the shape of the chaotic map by changing the feedback coefficient $\beta$. Fig. S2 shows the nonlinear maps when $\beta$ is changed. Figs. S2(a), S2(b), and S2(c) show the nonlinear maps for $\beta$ = 0.5, 1.0, and 2.0, respectively. A cosine function with one period ($2\pi$) is obtained for $\beta = 1.0$, and a cosine curve with two periods ($4\pi$) is obtained for $\beta = 2.0$. Therefore, the number of periods of the nonlinear function is controlled by $\beta$, and the number of peaks of the sinusoidal function increases with an increase in $\beta$.

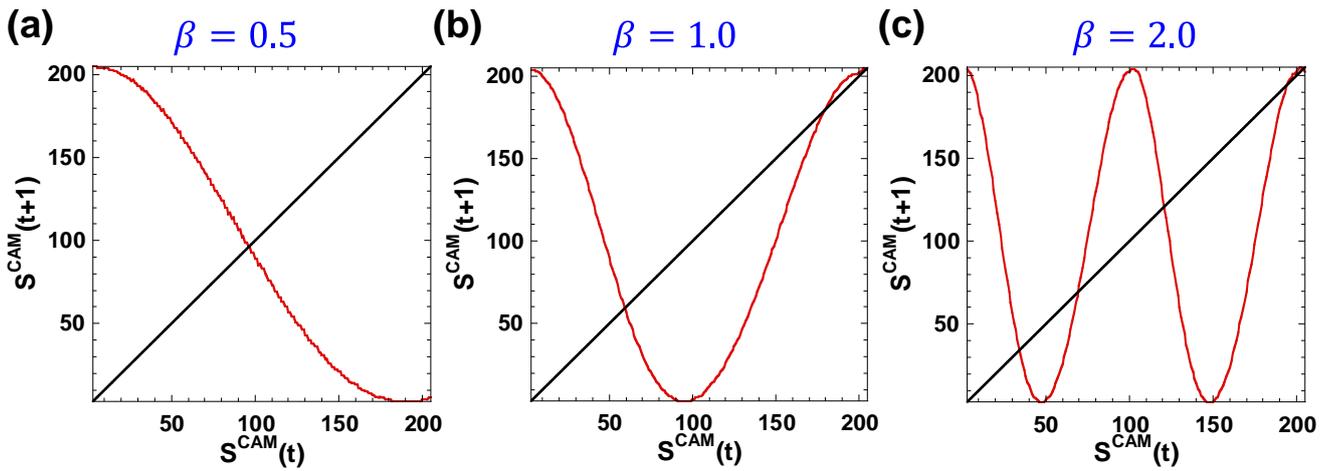

Figure S2. Nonlinear functions for the feedback coefficients (a) $\beta = 0.5$, (b) $\beta = 1.0$, and (c) $\beta = 2.0$.

Fig. S3 shows the bifurcation diagram and Lyapunov exponent of the chaotic map by changing $\beta$. Steady-state and periodic oscillations appear for small $\beta$. Chaotic oscillations are observed in the black regions of the bifurcation diagram. The regions for positive Lyapunov exponents correspond to the black regions of chaos in the bifurcation diagram. Therefore, we confirm the appearance of chaotic behavior by both the bifurcation diagram and Lyapunov exponent in numerical simulations.

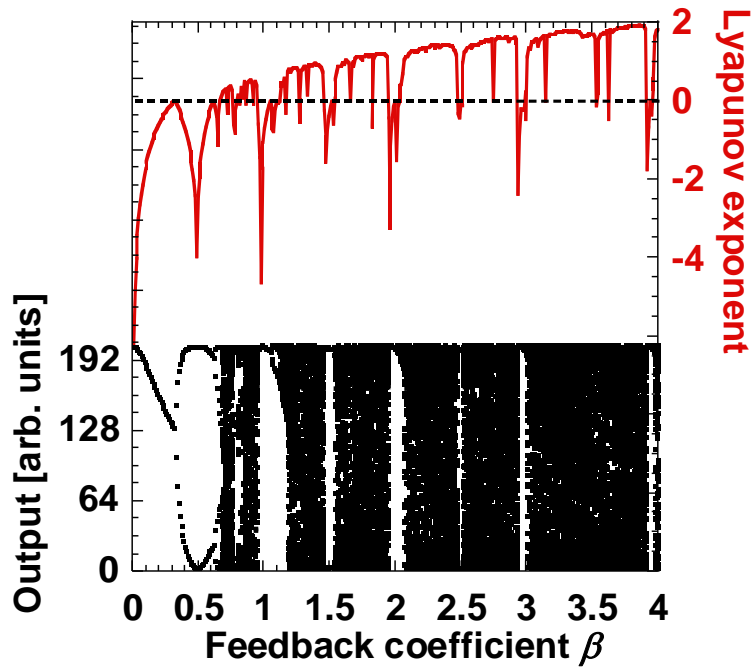

Figure S3. Numerical results of the bifurcation diagram (black, lower) and Lyapunov exponent (red, upper) as a function of feedback coefficient $\beta$.

*Derivation of the nonlinear map in the optoelectronic feedback system*

The SLM provides phase modulation in one polarization direction, which is converted into intensity modulation by a polarizer. The nonlinear function between the input and output intensities was experimentally observed at one of the macro-pixels on the SLM, as shown in Fig. S4(a). The nonlinear function of the SLM and the camera can be approximated as follows:

$$S_i^{CAM}(t) = a \cdot \cos(2\pi f(S_i^{SLM}(t) - \phi)) + b, \tag{S1}$$

where $S_i^{CAM}(t)$ and $S_i^{SLM}(t)$ are the optical intensities of macro-pixel $i$ at time $t$ on the camera and SLM, respectively. $S_i^{CAM}(t)$ and $S_i^{SLM}(t)$ are 8-bit integer signals ranging from 0 to 255. $a, b, f$, and $\phi$ represent the amplitude, offset, frequency, and phase of the nonlinear function, respectively (see Fig. S4(a)). The input signal to the SLM $S_i^{SLM}(t)$ is converted into the output signal at the camera $S_i^{CAM}(t)$ by magnifying the 2D pattern with the ratio between the whole pixel size of the SLM and camera. When generating the feedback signal, we compensate for the phase shift $\phi$ using $\acute{S}_i^{SLM}(t) = S_i^{SLM}(t) - \phi$, where $\phi = 23$ in Fig. S4(a). The nonlinear function in Fig. S4(a) is truncated to use a half period of the sinusoidal function (the blue arrow in Fig. S4(a)). Fig. S4(b) shows the half period of the sinusoidal function extracted from Fig. S4(a). This function forms the basis of optoelectronic feedback systems.

The feedback signal of the phase modulation to the SLM is manipulated in the PC and fed back to the SLM using the following formula:

$$\acute{S}_i^{SLM}(t+1) = \beta S_i^{CAM}(t) \, \% \, \alpha, \tag{S2}$$

where $\alpha$ is the effective range that determines the range of the nonlinear function corresponding to one period of a sinusoidal function in Fig. S4(b). We set $\alpha = 201$, which corresponds to the size of one period in Fig. S4(b). Here, % is the modulo operation, and $\beta$ is the feedback coefficient that determines the shape of the nonlinear function. The introduction of $\alpha$ and $\beta$ results in a periodic function. From Eqs. (S1) and (S2), a one-dimensional map can be derived as follows:

$$S_i^{CAM}(t+1) = a \cdot \cos(2\pi f \beta S_i^{CAM}(t) \, \% \, \alpha) + b, \tag{S3}$$

where we set $f = 1/\alpha$ so that % $\alpha$ can be omitted in Eq. (S3). Then, Eq. (S3) is equivalent to Eq. (1) in the main text.

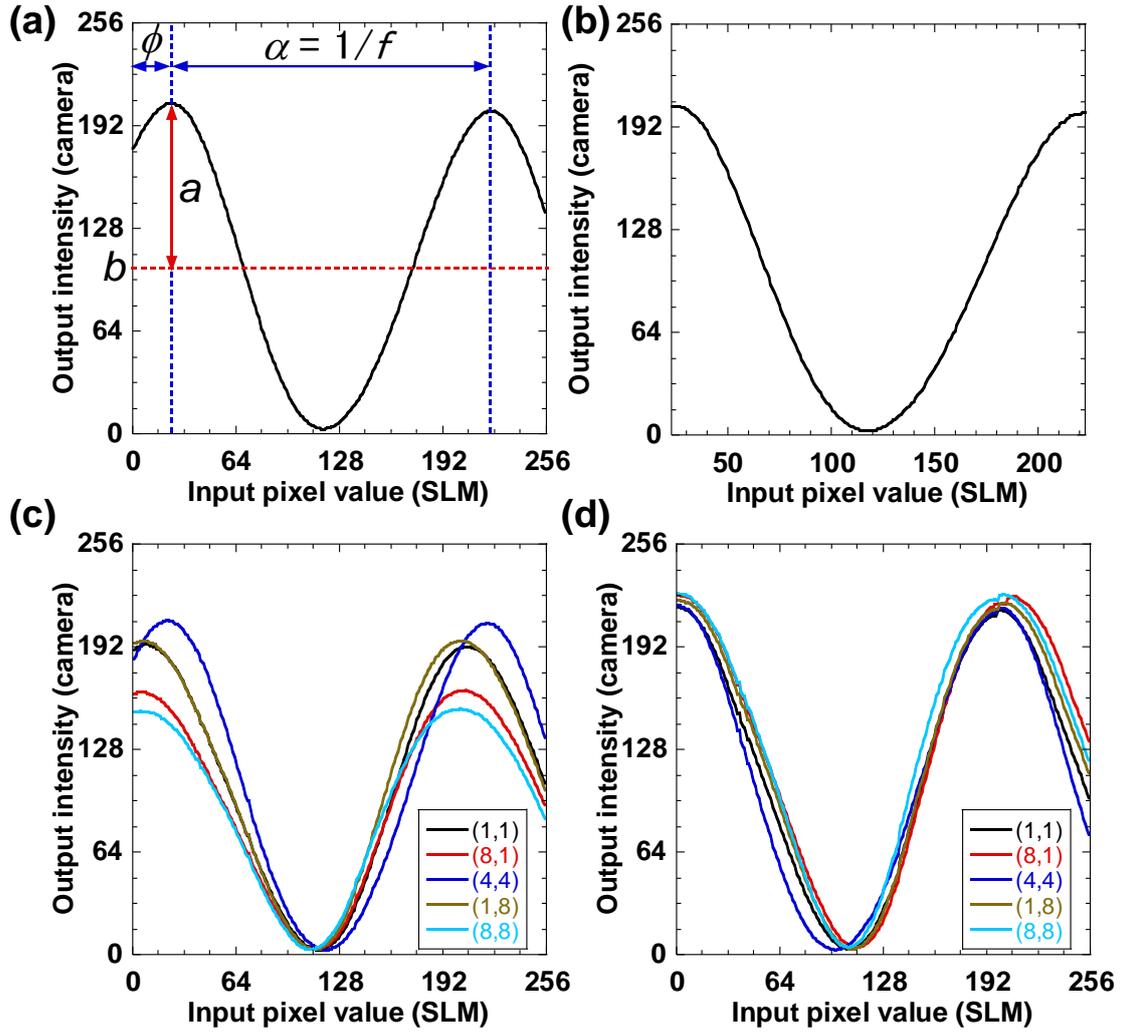

Figure S4. Experimentally obtained nonlinear functions of the optoelectronic feedback system. (a) Raw data of the nonlinear function between the optical input intensity on the SLM and the output intensity detected by the camera at the (4, 4) macro-pixel. (b) Truncated nonlinear function with one period of a sinusoidal function. (c) Raw data of nonlinear functions at different macro-pixels ($r$, $s$) at the $r$-th row and $s$-th column. (b) Adjustment of nonlinear functions obtained by changing the parameter values.

## Alignment of nonlinear functions for different macro-pixels

The shapes of the nonlinear functions are different for different macro-pixels owing to the inhomogeneity of the phase modulation characteristics on the SLM. The parameter values of the nonlinear functions were measured in the experiment and adjusted to be as similar as possible among the different macro-pixels. Fig. S4(c) shows the nonlinear functions of different macro-pixels ($r$, $s$) that are $8 \times 8$ in size. We matched the nonlinear functions for all the macro-pixels as closely as possible by adjusting the values of $a$, $b$, $\phi$, $\alpha$, and $f$. Fig. S4(d) shows the modified nonlinear functions for the different macro-pixels after parameter adjustment. The peaks of the nonlinear functions are matched, and the nonlinear functions overlap.

**Experimental result of decision making for 256-armed bandit problem**

We performed decision making to solve the bandit problem using 256 slot machines. Fig. S5 shows an example of the spatiotemporal pattern with biases for the decision-making process using 256 slot machines (16 × 16 macro-pixels). The hit probabilities are set as follows: $P_1 = 0.7$, $P_2 = 0.5$, $P_3 = 0.9$, $P_4 = 0.1$, …, $P_{2j-1} = 0.7$, and $P_{2j} = 0.5$, $j \geq 3$, where $j$ is an integer. Slot machine 3 is selected at the 4500$^{th}$ play, as indicated by the white pixel (Fig. S5(b)), indicating that correct decision making can be achieved.

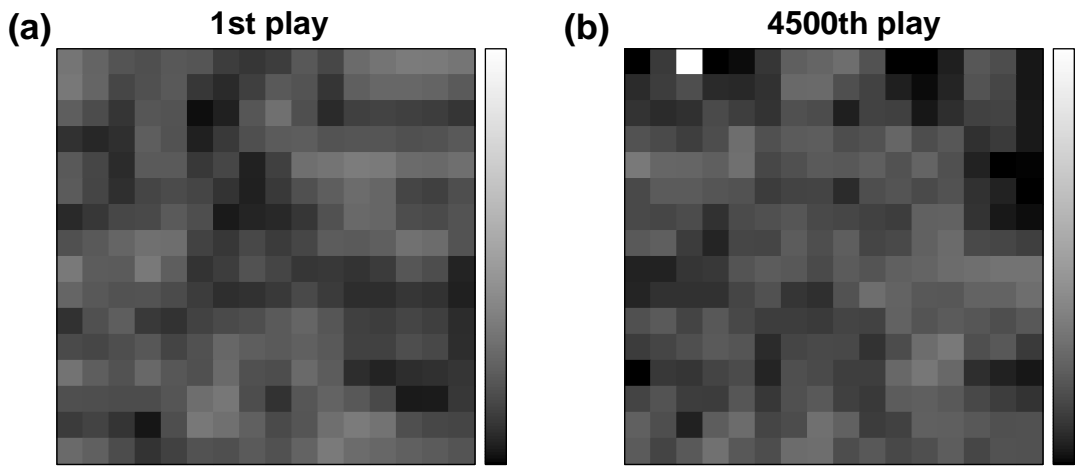

Figure S5. Decision-making process for solving the multi-armed bandit problem with 256 slot machines at (a) the first and (b) the final (4500$^{th}$) plays. White macro-pixel at (1, 3) in (b) indicates the final decision making (slot machine 3). 16 × 16 macro-pixels are used.

**Movies illustrating the spatiotemporal dynamics and decision-making process**

Supplementary movies are provided to observe the spatiotemporal dynamics and decision-making process for solving the 64-armed bandit problem observed in experiment.

Movie S1. Spatiotemporal dynamics of an optoelectronic feedback system measured by a CMOS camera in experiment. 8 × 8 macro-pixels are used. White and black color correspond to high and low optical intensity, respectively, as shown in Fig. 2(a) in the main text.

Movie S2. Decision-making process for solving the 64-armed bandit problem. 8 × 8 macro-pixels are assigned to 64 slot machines. White macro-pixel at (1, 3) indicates the final decision of the slot machine with the maximum hit probability (slot machine 3).